\documentclass{article}
\usepackage{graphicx}
\title{\bf Spinning particles in Schwarzschild-de Sitter space-time}
\author{M. Mortazavimanesh and Morteza Mohseni\thanks{email:
m-mohseni@pnu.ac.ir} \\{\small Physics Department, Payame Noor University, Tehran 19395-4697,
Iran}}
\begin{document}
\maketitle
\begin{abstract}
After considering the reference case of the motion of spinning test bodies in the equatorial plane of the Schwarzschild space-time, we
generalize the results to the case of the motion of a spinning particle in the equatorial plane of the Schwarzschild-de Sitter space-time.
Specifically, we obtain the loci of turning points of the particle in this plane. We show that the cosmological constant affect the particle motion when the particle distance from the black hole is of the order of the inverse square root of the cosmological constant.

Keywords: Spinning particles, Schwarzschild-de Sitter space-time

PACS: 04.70.Bw; 98.80.Es
\end{abstract}
\section{Introduction}
The dynamics of particles in the gravitational field of a black hole is an interesting problem and has been studied extensively in the past
(see e.g. \cite{chan}). When the test particle has some internal angular momentum (spin), one could expect extra effects due to the interaction of the particle spin with the gravitational field. The well known Mathisson-Papapetrou-Dixon (MPD) equations describe the motion of such bodies in a general curved background by taking the spin-curvature interaction into account \cite{dix}.

The motion of a spinning particle in the space-time of a Kerr black hole has been studied in \cite{tod}. That study resulted in an exact solution for the MPD equations in the special case of motion in the equatorial plane with the particle spin perpendicular to this plane. A similar problem, the motion in the equatorial plane but with varying spin components was investigated in \cite{bini}. The motion of spinning particles in a Kerr field has also been studied in \cite{sem1,sem2} using numerical techniques and various supplementary conditions. Spatially circular orbits of spinning particle in Schwarzschild space-time and the resulting clock effects were studied in \cite{bini2} and ultra-relativistic circular orbits were studied in \cite{ply}. The clock effects were also investigated for the Kerr black hole in \cite{gera}. The case of the motion in Reissner-Nordstr\"{o}me space-time was studied in \cite{jant} and in \cite{yuan}.

The aim of the present work is to investigate the effect of the cosmological constant on the dynamics of spinning particles. To this end, we
consider the motion of such particles in a Schwarzschild-de Sitter space-time. The cosmological constant has been a subject of interest since the early stages of the development of the general theory of relativity. This interest has been revived by connecting it with the vacuum energy of a quantum field \cite{win}. It has also been echoed recently by the extensive research on dark energy for which the cosmological constant is the simplest candidate (see e.g \cite{pad}).

The role of cosmological constant in the dynamics of particles has been investigated in the literature in different contexts.
The motion of test particles in Schwarzschild-de Sitter and Schwarzschild-anti de Sitter space-times has been studied in \cite{hle2}. Equilibrium of spinning test particles in Schwarzschild-de Sitter space-time has been investigated in \cite{stu1} in which it has been shown that the relevant condition (which is independent of the spin of the particle) is met only on the so called static radius where the gravitational attraction is compensated by the repulsive cosmological constant. The case of charged spinning particles in Reissner-Nordstr\"{o}m space-time has been considered in \cite{hle1} and motion in Reissner-Nordstr\"{o}m-de Sitter space-time has been studied in \cite{ali}. Equilibrium conditions for spinning particles in a Kerr-de Sitter space-time was considered in \cite{hor}. A relativistic top theory has been discussed in \cite{ni1} from the point of view of the de Sitter group and some other mathematical aspects of this theory has been studied in \cite{ni2}.

In the following sections we first review the MPD equations and subsequently, as a reference case, we consider the motion of a spinning particle in the equatorial plane of the Schwarzschild space-time. The results are in agreement with those of the more general case of the Kerr black hole \cite{tod} in the appropriate limit. We then generalize the results to the case of the Schwarzschild-de Sitter space-time. The cosmological constant affects the dynamics of the particle in distances far from the black hole. These distances are of the order of inverse square root of the cosmological constant. In the last section we discuss the results.
\section{The Equations of Motion}
The motion of a spinning particle moving in a curved space-time is described by the MPD equations. To the lowest order, these are given by \cite{dix}
\begin{eqnarray}
{\dot s^{\mu\nu}}&=&p^\mu v^\nu-p^\nu v^\mu,\label{eq3a}\\
{\dot p^\mu}&=&-\frac{1}{2}{R^\mu}_{\nu\alpha\beta}v^\nu
s^{\alpha\beta},\label{eq3b}
\end{eqnarray}
where $s^{\mu\nu}$ is the particle spin tensor, $p^\mu$ is the particle momentum, $R_{\mu\nu\kappa\lambda}$ is the curvature tensor, overdots mean covariant differentiation $v^\alpha\nabla_\alpha$, and $v^\mu$ is the particle velocity. These equations are usually supplemented by another equation to make them complete. A well-known example is the Tulczyjew supplementary condition
\begin{equation}\label{eq4}
p_\mu s^{\mu\nu}=0.
\end{equation}
It can be shown that under these equations the particle mass and spin are conserved, i.e.,
\begin{eqnarray}
p_\mu p^\mu&=&\mbox{const.}=-m^2,\label{h1}\\
s_{\mu\nu}s^{\mu\nu}&=&\mbox{const.}=2s^2.\label{h2}
\end{eqnarray}
It is convenient to represent the particle spin with the spin four-vector defined by
\begin{equation}\label{m1}
s^\mu=\frac{1}{2m\sqrt{-g}}\epsilon^{\mu\nu\alpha\beta}p_\nu s_{\alpha\beta}
\end{equation}
in which $\epsilon^{\mu\nu\alpha\beta}$ is the alternating symbol. There is no equation of motion for the particle velocity but it can be computed
via the following equation
\begin{equation}\label{eq11}
v^\mu=\frac{v_\kappa p^\kappa}{p_\lambda p^\lambda}\left(p^\mu+\frac{2s^{\mu\nu}R_{\nu\rho\alpha\beta}
s^{\alpha\beta}p^\rho}{-4p_\gamma p^\gamma+s^{\delta\eta}R_{\delta\eta\epsilon\zeta}s^{\epsilon\zeta}}\right)
\end{equation}
which is itself a consequence of the equations of motion \cite{tod,ehl}. This equation simplifies further when some specific gauges, say
$v_\kappa p^\kappa=-m$, is chosen. This is the gauge in which the instantaneous zero-momentum and the zero-velocity frames are simultaneous \cite{dix}. The MPD equations do not guarantee that $v^\mu v_\mu$ is constant. Denoting $\frac{1}{m}p^\mu=u^\mu$, the vector $f^\mu=v^\mu-u^\mu$ is orthogonal to $u^\mu$ and is the particle 3-velocity with respect to the zero-momentum observer with 4-velocity $u^\mu$ \cite{ehl}.

For a space-time admitting a Killing vector $\xi^\mu$ the following relation gives a constant of motion \cite{dix}
\begin{equation}\label{e5}
p^\mu\xi_\mu+\frac{1}{4}(\nabla_\nu\xi_\mu-\nabla_\mu\xi_\nu)s^{\mu\nu}=\mbox{cons.}
\end{equation}
\section{Motion in Schwarzschild Space-time}
We first consider the motion of a spinning particle in the equatorial plane of the Schwarzschild black hole with the particle spin vector perpendicular to that plane. This problem has been solved in \cite{tod} for the more general case of a Kerr black hole. The Schwarzschild metric is a spherically symmetric solution of the vacuum Einstein field equations. In terms of a spherical coordinates $x^\mu=(t,r,\theta,\phi)$, it is given by
\begin{equation}\label{m2}
ds^2=-f(r)dt^2+\frac{1}{f(r)}dr^2+r^2d\theta^2+r^2\sin^2\theta d\phi^2
\end{equation}
where $f(r)=1-\frac{2M}{r}$, $M$ being the gravitational source.

We now take the spin vector $s^\mu$ in the $\theta$-direction with
\begin{equation}\label{e51}
s^\mu=\frac{s}{r}(0,0,1,0),
\end{equation}
in which $s$ is constant. The particle four-momentum is perpendicular to spin by construction, equation (\ref{m1}). Here we adopt $\epsilon^{tr\theta\phi}=+1$. The above choice for the spin vector is consistent with the equations of motion (\ref{eq3a}) and (\ref{eq3b}) (see also \cite{hojh}).

Now by starting from equation (\ref{eq11}), it can be shown that the following relation holds
\begin{equation}\label{eq11a}
v^r=\frac{-v_\alpha p^\alpha}{m^2}\left(1+\frac{3M{\hat s}^2(p^\phi)^2r^2}{m^2r^3+{\hat s}^2M(3(p^\phi)^2-m^2)}\right)p^r
\end{equation}
in which ${\hat s}=\frac{s}{m}$. This shows that $p^r$ and $v^r$ are proportional if the factor in the right hand side of the above relation is not equal to zero. Thus the zeros of $p^r$ and $v^r$ coincide. Now from (\ref{h1}) we have
\begin{equation}
-m^2=g^{tt}(p_t)^2+g^{rr}(p_r)^2+g^{\theta\theta}(p_\theta)^2+g^{\phi\phi}(p_\phi)^2.
\end{equation}
Since we are interested in turning points, we set $p_r=0$. Also according to the chosen ansatz, the particle has a vanishing  $p_\theta$ component. Thus the second and third term in the right hand side of the above expression drop and we are left with
\begin{equation}\label{x1}
-m^2=g^{tt}(p_t)^2+g^{\phi\phi}(p_\phi)^2.
\end{equation}
To obtain the components $p_t$ and $p_\phi$ of the momentum we use equation (\ref{e5}). The Schwarzschild metric admit four Killing vectors from which we choose the following ones
\begin{eqnarray}
\xi^\mu_t&=&(1,0,0,0),\label{d1}\\
\xi^\mu_\phi&=&(0,0,0,1)\label{d2}
\end{eqnarray}
representing its stationary and axisymmetric nature. As a consequence, equation (\ref{e5}) gives the following constants of motion
\begin{eqnarray}
p_t+\frac{1}{2}\partial_\nu g_{0\mu}s^{\mu\nu}&=&-E,\label{d3}\\
p_\phi+\frac{1}{2}\partial_\nu g_{3\mu}s^{\mu\nu}&=&l\label{d4}
\end{eqnarray}
in which $E,l$ can be interpreted as the particle energy and the angular momentum component along the axis of symmetry, respectively.

Thus equations (\ref{d3}) and (\ref{d4}) can be recast in the following form
\begin{eqnarray}
p_t+p_\phi{\hat s}\frac{M}{r^3}&=&-E,\label{d3a}\\
p_\phi+p_t{\hat s}&=&l\label{d4a}.
\end{eqnarray}
Defining $A=\frac{M\hat s}{r^3}$, from these equations we obtain
\begin{eqnarray}
p_t&=&(Al+E)(A{\hat s}-1)^{-1},\label{d3b}\\
p_\phi&=&-(l+E{\hat s})(A{\hat s}-1)^{-1}.\label{d4b}
\end{eqnarray}
Putting these into (\ref{x1}) and solving the resulting equation for ${\cal E}=\frac{E}{m}$ we obtain
\begin{equation}\label{e51ha}
{\cal E}_\pm=\frac{-\lambda(f{\hat s}-r^2A)\pm r(1-A{\hat s})\sqrt{f(\lambda^2-f{\hat s}^2+r^2)}}{f{\hat s}^2-r^2}
\end{equation}
where $\lambda=\frac{l}{m}$, and $f=1-\frac{2M}{r}$. This result agrees with equation (38) of \cite{tod} in the limiting case of $a=0$.

\begin{figure}[h]
\includegraphics[height=4.1cm,width=6cm,angle=-90]{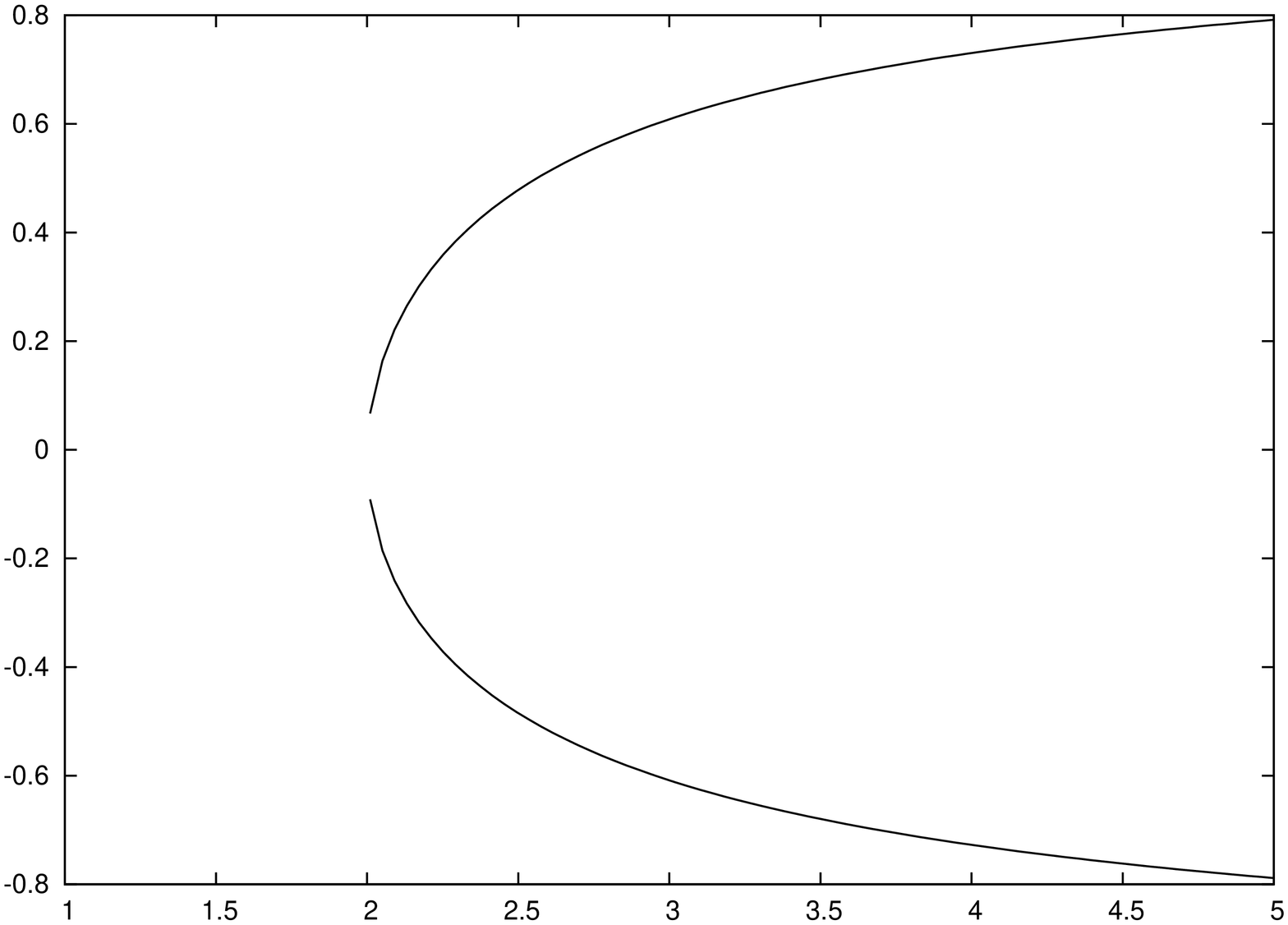}\hspace{2cm}\includegraphics[height=4.1cm,width=6cm,angle=-90]{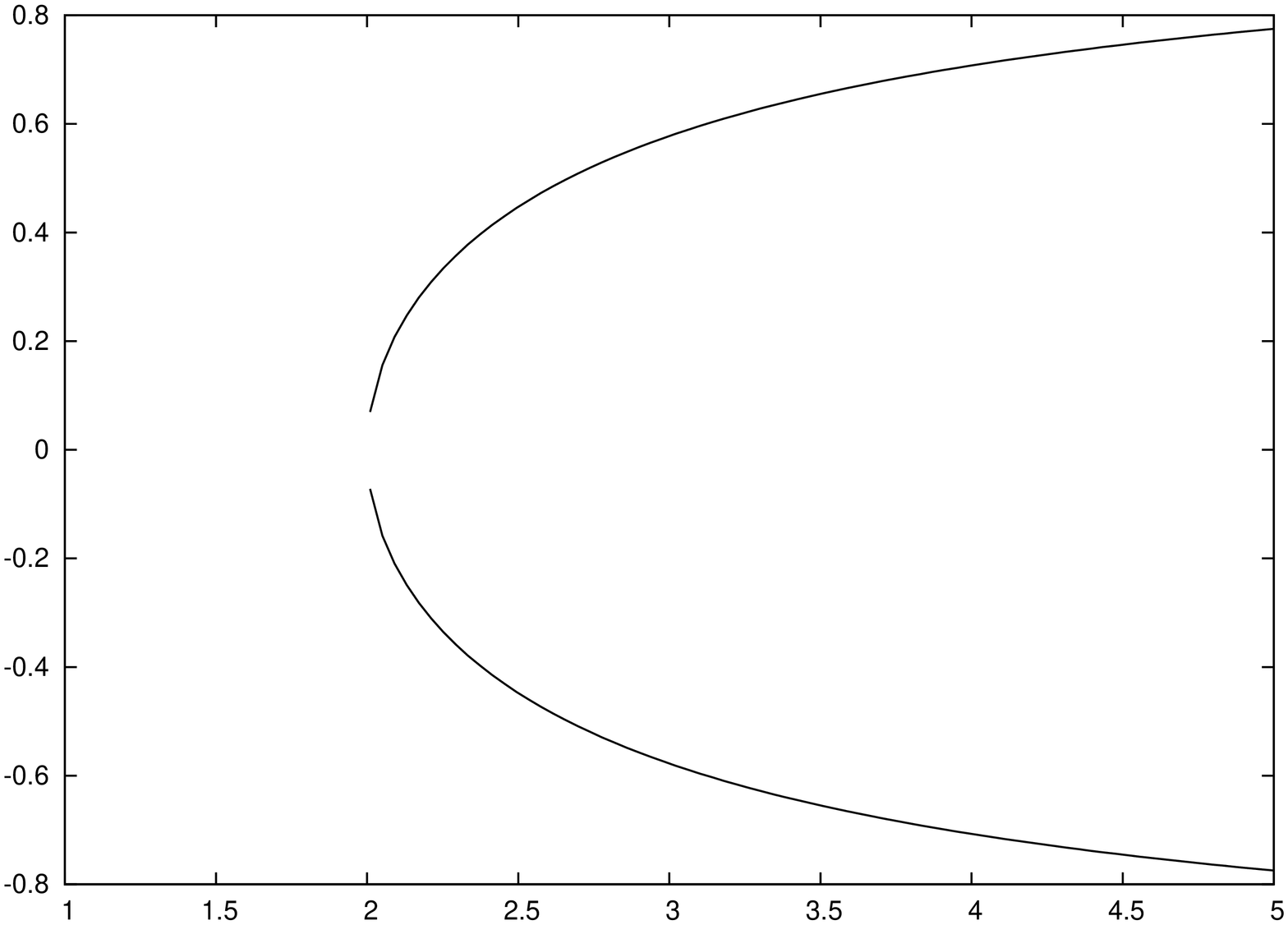}\vspace{-1.5cm}
\caption{The loci of turning points for the motion in Schwarzschild space-time for $M=1$, $\lambda=1$, ${\hat s}=0.1$ (left), and $\lambda=0.1,{\hat s}=0.1$ (right). Lower branches correspond to the particle specific energy ${\cal E}_+$ and upper branches to ${\cal E}_-$. The $E$-axis (vertical axis) is in units of $m$.}
\label{fi4a}
\end{figure}

The validity of the Mathisson-Papapetrou-Dixon model for spinning test particles is subject to the condition that the characteristic length scale $\hat s$
associated with the particle be small compared with the natural length scale $M$, associated with the background field \cite{mol}. We have depicted ${\cal E}_\pm$ in figure \ref{fi4a} which shows the loci of turning points of the spinning particle for different values of ${\hat s}$ and $\lambda$.

\begin{figure}[h]
\begin{center}
\includegraphics[height=5cm,angle=-90]{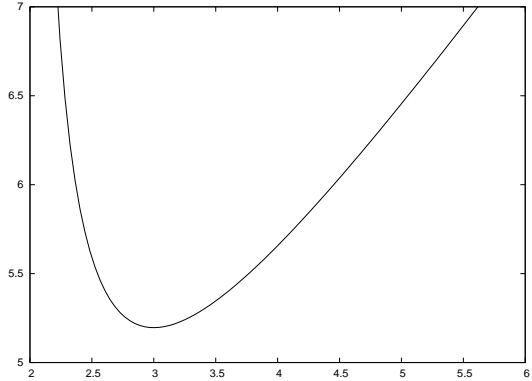}\vspace{-1.5cm}
\caption{The particle spin for the Schwarzschild space-time. The $r$-axis (horizontal axis) is in units of $M$ and the $s$-axis is in units of $mM$.}
\label{fig1}\end{center}
\end{figure}

For $f{\hat s}^2-r^2=0$ the function ${\cal E}$ diverges. This situation is equivalent to setting $${\hat s}=\pm r\sqrt{\frac{r}{r-2M}}.$$ This is depicted in figure \ref{fig1}. The figure shows that ${\hat s}$ has a minimum value of about $5M$ which is far beyond the values for which the whole approach is valid. This reaches us at the conclusion that ${\cal E}_\pm$ do not diverge within this model.

We now consider the case where $v^r$ is zero for non-zero $p^r$. From equation (\ref{eq11}), this corresponds to
\begin{eqnarray*}
1+\frac{3M{\hat s}^2(p^\phi)^2r^2}{m^2r^3+{\hat s}^2M(3(p^\phi)^2-m^2)}=0
\end{eqnarray*}
or
\begin{equation}\label{o1}
p^\phi=\pm\frac{m}{\hat s\sqrt{3M}}\sqrt{\frac{M{\hat s}^2-r^3}{1+r^2}}
\end{equation}
which is meaningless because for small ${\hat s}$ the numerator $M{\hat s}^2-r^3$ is negative.
The occurrence of this case might be a hint at the limit of the validity of the method used here. Thus this case is ruled out.

For small $\hat s$ one can expand expression (\ref{e51ha}) in powers of $\hat s$. Expanding (\ref{e51ha}) and neglecting $O({\hat s}^2)$ terms, we
get the following relation for the particle specific energy
\begin{equation}\label{e51hb}
{\cal E}_{\pm}=\frac{\mp\sqrt{f(\lambda^2+r^2)}}{r}-\frac{\lambda{\hat s}M}{r^3}
\end{equation}
which is more illuminating.
\section{Motion in Schwarzschild-de Sitter space-time}
The static spherically symmetric solution of the Einstein field equations with cosmological constant term
\begin{equation}\label{y1}
G_{\mu\nu}-\Lambda g_{\mu\nu}=-\kappa T_{\mu\nu}
\end{equation}
for vanishing sources $T_{\mu\nu}=0$ is the Schwarzschild-de Sitter space-time solution which can be written in Schwarzschild-like coordinates as
\begin{eqnarray}\label{m2a}
ds^2=-\left(1-\frac{2M}{r}+\frac{\Lambda}{3}r^2\right)dt^2+\frac{dr^2}{1-\frac{2M}{r}+\frac{\Lambda}{3}r^2}
+r^2d\theta^2+r^2\sin^2\theta d\phi^2.
\end{eqnarray}
Taking equations (\ref{d3}) and (\ref{d4}) (with the same Killing vectors) into account, we find for the above space-time
\begin{eqnarray}
p_t+\left(\frac{M}{r^3}+\frac{\Lambda}{3}\right)p_\phi{\hat s}&=&-E,\label{d33}\\
p_\phi+p_t{\hat s}&=&l\label{d44}.
\end{eqnarray}
These result in
\begin{eqnarray}
p_t&=&(Bl+E)(B{\hat s}-1)^{-1},\label{d31}\\
p_\phi&=&-(l+E{\hat s})(B{\hat s}-1)^{-1}.\label{d42}
\end{eqnarray}
where $B=\frac{M{\hat s}}{r^3}+\frac{\Lambda}{3}{\hat s}.$ Now by following the same procedure as of the previous section, we reach at
\begin{equation}\label{e51h}
{\cal E}_\pm=\frac{-\lambda(h{\hat s}-r^2B)\pm r(1-B{\hat s})\sqrt{h(\lambda^2-h{\hat s}^2+r^2)}}{h{\hat s}^2-r^2}
\end{equation}
in which $h=1-\frac{2M}{r}+\frac{\Lambda}{3}r^2$.

For the region near the black hole, the particle behaves in the same manner as in the case of Schwarzschild space-time due to the smallness of the cosmological constant. Thus according to the above relation the cosmological constant has a significant effect only for large distances $r_0$ which are of the order of
\begin{equation}\label{ad1}
r_0\sim\frac{1}{\sqrt\Lambda}\sim\frac{1}{\sqrt{10^{-56}}}\sim 10^{28}cm.
\end{equation}
The value of the cosmological constant used above could be related to other physical parameters such as the critical density $\rho_c$ and the Hubble parameter $$H_0=100h\hspace{2mm}kms^{-1}Mpc^{-1}$$ with the dimensionless parameter $h\sim 0.7$ \cite{wilk}.

\begin{figure}[h]
\begin{center}
\includegraphics[height=5cm,angle=-90]{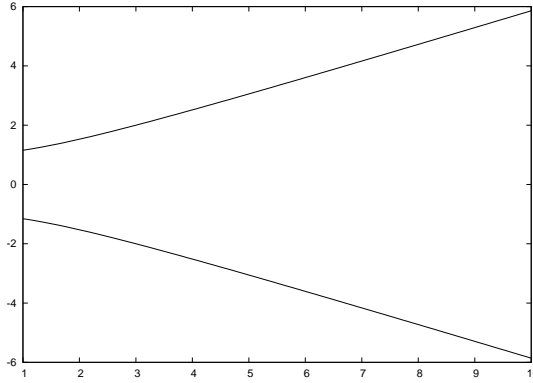}\vspace{-2cm}
\caption{The locus of turning points for the motion in Schwarzschild-de Sitter  space-time for $r\sim r_0$ Upper branch corresponds to the particle specific energy ${\cal E}_-$ and lower branch to ${\cal E}_+$. The $r$-axis is in units of $\frac{1}{\sqrt\Lambda}$ and the $E$-axis is in units of
$m$.}\label{fi5}\end{center}
\end{figure}

For distances of this order of magnitude, the particle energy becomes independent of its spin ${\hat s}$.
The above relation is depicted in figure \ref{fi5}.

In the limit $h{\hat s}^2-r^2=0$ which corresponds to
\begin{equation}\label{g43}
{\hat s}=\pm r^{\frac{3}{2}}\left(r-2M+\frac{\Lambda r^3}{3}\right)^{-\frac{1}{2}}
\end{equation}
the particle specific energy ${\cal E}$ diverges. This expression reduces to the corresponding relation for the case of Schwarzschild space-time whenever we set $\Lambda=0$. For values of $r$ comparable to $r_0$, the cosmological constant becomes important, but this corresponds to values of
${\hat s}$ which are not within the range of validity and thus we get the same conclusion as to the case of motion in Schwarzschild space-time.

The arguments related to the situation where $p^r$ is not equal to zero but $v^r$ vanishes go as follows. We have the relation
\begin{equation}\label{eq11b}
v^r=\frac{-v_\alpha p^\alpha}{m^2}\left(1+\frac{9M{\hat s}^2(p^\phi)^2r^2}{3m^2r^3+{\hat s}^2\left(3M(3(p^\phi)^2-m^2)-\Lambda m^2r^3\right)}\right)p^r
\end{equation}
and hence
\begin{equation}\label{o2}
p^\phi=\pm\frac{m}{3\hat s\sqrt{M}}\sqrt{\frac{3M{\hat s}^2+(\Lambda{\hat s}^2-3)r^3}{1+r^2}}
\end{equation}
resulting to the same conclusion as to the case described by relation (\ref{o1}).

For small values of particle specific spin ${\hat s}$ we can obtain a relation similar to (\ref{e51hb})
\begin{equation}\label{e51hc}
{\cal E}_{\pm}=\frac{\mp\sqrt{h(\lambda^2+r^2)}}{r}-\lambda{\hat s}\left(\frac{M}{r^3}+\frac{\Lambda}{3}\right).
\end{equation}

For $M=0$, the metric (\ref{m2a}) reduces to de Sitter metric. For this space-time, it can be shown from equation (\ref{eq11}) that $p^\mu$ is equal to $mv^\mu$, just like a test particle without spin. This result is also reflected in the expression for the particle specific energy. Setting $M=0$ in
(\ref{e51hc}) results in
\begin{equation}\label{e51hd}
{\cal E}_{\pm}=\frac{\mp\sqrt{h^\prime(\lambda^2+r^2)}}{r}-\frac{\lambda{\hat s}\Lambda}{3}
\end{equation}
where $h^\prime=1+\frac{\Lambda}{3}r^2$. This shows that the particle spin contribution to ${\cal E}$ is a constant term.

\begin{figure}[h]
\begin{center}
\includegraphics[height=5cm,angle=-90]{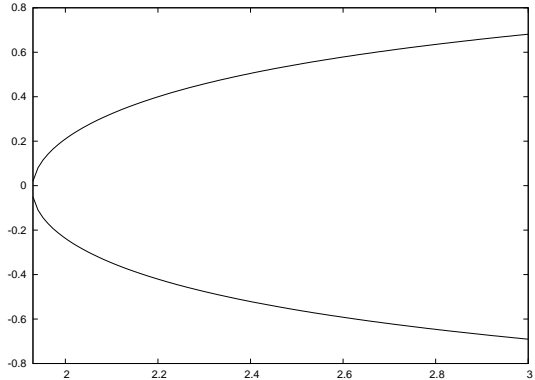}\vspace{-2cm}
\caption{The locus of turning points for the motion in Schwarzschild-de Sitter  space-time for $M=1,\lambda=1,{\hat s}=0.1, \Lambda=0.03$.  Upper branch corresponds to the particle specific energy ${\cal E}_-$ and lower branch to ${\cal E}_+$. The vertical axis is in units of
$m$.}\label{fi0}\end{center}
\end{figure}

So far in our analysis, we have confined ourselves to the present time value of the cosmological constant which is very small. However it would also be interesting to consider an effective cosmological constant of higher orders of magnitude corresponding to some other epoches. In fact, substantially higher orders of magnitude could have been occurred in the early universe due to phase transitions \cite{refer}. This leads to a different behavior
of spinning particles. Figure \ref{fi0} shows the particle specific energy in such circumstances.
\section{Discussion}
We have studied the motion of a spinning particle in the equatorial plane of the Schwarzschild and Schwarzschild-de Sitter space-times
with the particle spin perpendicular to that plane. Our analysis is based on a systematic use of both generic and non-generic (i.e those depending on the background symmetries) constants of motion. For the Schwarzschild case the corresponding results of the case of the Kerr black hole \cite{tod}
are recovered in the appropriate limit, up to differences due to imposing the validity condition ${\hat s}\ll M$.

In both space-times, the turning points are located closer to the black hole for lower particle energies. In the case of the Schwarzschild-de Sitter space-time the particle dynamics is affected by the cosmological constants only for distances of the order of inverse square root of the cosmological constant. For these distances the loci of turning points have a different shape compared to distances close to the black hole.

For vanishing $v^\phi$ the locus of turning points we have found, is the equilibrium position of spinning particle
in the equatorial plane. Thus our results are related to those of \cite{stu1,hle1} in which the equilibrium position have been obtained (but with a
different supplementary condition.) In \cite{hle1}, the authors obtained the result that in the space-time under consideration, the equation of motion in not influenced by the spin vector in the case of particles in stationary equilibrium positions. Concerning the motion of particles, our considerations are more general, because we have not assumed the stationary equilibrium condition. For $\hat s=0$, parts of the results of \cite{hle2} are recovered.

We have used the same supplementary condition as that of  \cite{tod}, however it would be interesting to consider other supplementary conditions as discussed in \cite{sem2}. Our results might be of interest in understanding the influence of Schwarzschild- de Sitter black hole and the cosmological constant on the behavior of spinning compact objects.

{\bf acknowledgements.} The authors would like to thank H.H. von Borzeskowski for useful comments.

\end{document}